\relax
\documentclass[fleqn,10pt]{wlscirep} 
\usepackage[utf8]{inputenc}
\usepackage[T1]{fontenc}
\usepackage{times}  
\usepackage{helvet}  
\usepackage{courier}  
\usepackage{url}  
\usepackage{graphicx}  

\usepackage{authblk} 

\usepackage{xcolor}
\usepackage{subfig}
\usepackage{graphicx}
\usepackage{graphics}
\usepackage{multirow}
\usepackage{nameref}

\usepackage{url}
\usepackage{times}
\usepackage{comment}
\usepackage{soul}

\usepackage[symbol]{footmisc}

\usepackage{soul}

\usepackage{todonotes}

\makeatletter
\@addtoreset{subfigure}{row}
\makeatother

\title{Analysing Meso and Macro conversation structures in an online suicide support forum}

\author[1,*]{Sagar Joglekar}
\author[1,2]{Sumithra Velupillai}
\author[2,]{Rina Dutta}
\author[1]{Nishanth Sastry}
\affil[1]{King's College, Department of Informatics, London, UK}
\affil[2]{King’s College London, IoPPN, London, SE5 8AF, UK}

\affil[*]{sagar.joglekar@kcl.ac.uk}

\keywords{Metal health, Network structure, Triadic motifs, Reddit}

\begin{abstract}
Platforms like Reddit and Twitter offer internet users an opportunity to talk about diverse issues, including those pertaining to physical and mental health. 
Some of these forums also function as a safe space for severely distressed mental health patients to get social support from peers. The online community platform Reddit's SuicideWatch is one example of an online forum dedicated specifically to people who suffer from suicidal thoughts, or who are concerned about people who might be at risk. It remains to be seen if these forums can be used to understand and model the nature of online social support, not least because of the noisy and informal nature of conversations. Moreover, understanding how a community of volunteering peers react to calls for help in cases of suicidal posts, would help to
devise better tools for online mitigation of such episodes. In this paper, we propose an approach to characterise conversations in online forums. Using data from the SuicideWatch subreddit as a case study, we propose metrics at a macroscopic level -- measuring the structure of the entire conversation as a whole. We also develop a framework to measure structures in supportive conversations at a mesoscopic level
-- measuring interactions with the immediate neighbours of the person in distress. We statistically show through comparison with baseline conversations from random Reddit threads that certain macro and meso-scale structures in an online conversation  exhibit signatures of social support, and are particularly over-expressed in SuicideWatch conversations. 
\end{abstract}
\begin{document}
\flushbottom
\maketitle

\vspace{-8mm}
\section{Introduction}
Suicide is responsible for 1·5\% of global mortality and is one of the most challenging public mental health 
issues \cite{oconnor_psychology_2014}.
Suicidality includes any thoughts or actions by an individual that could 
result in death \cite{turecki_suicide_2016}. 
Preventing death by suicide is a priority for health care services internationally \cite{zalsman_suicide_2016}, but poses a great challenge since accurately predicting an episode of suicidality is almost impossible\cite{mchugh2019,velupillai_risk_2019}. Furthermore, many deaths by suicide occur in people who did not have a known diagnosed mental health condition when they died\cite{stone_vital_2018}. Our ability to understand suicide has therefore been hampered by our ability to obtain data ``in situ''~\cite{nock2019}.
Platforms such as Reddit and Twitter are starting to offer a new and uniquely transparent window into suicide and other mental health issues. Unlike traditional health records, social media posts are authored by the users themselves. Also, in contrast to formal clinical settings, users on such platforms express themselves freely rather than regulating answers to establish a positive impression or be socially desirable \cite{van_de_mortel_role_2008}, thus providing a fresh and honest perspective~\cite{gkotsis2017characterisation}. Social media have therefore become a fertile ground for mental health studies, leading to new results in 
depression, anxiety, autism, and other problems~\cite{de2014mental,park2018examining,shen2017detecting,park2017longitudinal,DeChoudhury2016,DeChoudhury2014}.
 
Recent studies have shown promising results in modeling and measuring signals and patterns in Reddit communities related to mental health. For instance, statistical relations of mental health and depression communities with suicide ideation have been studied \cite{DeChoudhury2014,DeChoudhury2016}. The authors explored linguistic and social characteristics that evaluate users' propensity to suicidal ideation. Approaches to classify reddit posts as related to certain mental health conditions have also been successfully developed, showing that there are certain characteristics specific to mental health-related topics in posts that can be automatically captured\cite{gkotsis2017characterisation}. Furthermore, in a study focused on reddit posts related to anxiety, depression and post-traumatic stress disorder, the authors show that these online communities exhibit themes of a supportive nature, e.g. gratitude for receiving emotional support\cite{park2018examining}. Positive effects of  participation in such fora have also been shown by improvements in members' written communication\cite{info:doi/10.2196/jmir.8219}. The supportive nature of comments in the SuicideWatch forum has also been studied by automatic identification and classification of helpful comments with promising results\cite{Kavuluru:2016:CHC:2975167.2975170}.
Naturally, several studies that have been based on these types of online communities 
look at the \emph{textual} content of these online fora and produce inferences about psychological states. In our work, we conjecture that apart from textual metrics, it is important to quantify the differences in the \emph{structure} of a supportive conversation.

In the context of suicide, social media occupies an important ``clinical whitespace''~\cite{coppersmith2018} --  long intervals between clinical encounters that are filled with frequent posts on social media. These provide the potential for increased visibility into patient mental states. Studies have started to use social media posts both to understand population level responses to external triggers such as celebrity suicides~\cite{kumardetecting,karamshuk} and as a means to assess suicide risk~\cite{shing2018,clpsych}.   

While these are important first steps, such studies 
could be affected by the very nature of social media: it is unregulated, and problems such as cyberbullying\cite{luxton2012social,patton2014social} could potentially affect the variables being studied.  For instance, risk of suicide may be exacerbated just by participating on online social platforms. Therefore, we believe it is crucial to understand first the \emph{nature} of the conversations that happen online around suicide and suicidal behaviour. This paper aims to answer the question:  Does social media activity provide a supportive medium for potentially vulnerable patients? We address this  by studying the interactions of users on  \textbf{SuicideWatch} (\url{https://www.reddit.com/r/SuicideWatch/}), an online community (``subreddit'') on the social media platform Reddit. SuicideWatch is a heavily moderated forum, keeping messages and conversations on topic, and is  focussed solely on the topic of suicide. 
The moderators take the message of peer support seriously, and are governed by guidelines that prohibits false promises, abuse, tough love and other clinically concerning methods of conversations~\cite{BBC_suicidewatch}
It has therefore been the focus of recent research~\cite{shing2018} and shared tasks which aim to advance the state of the art~\cite{clpsych}. 

In contrast with previous studies which only looked at original posts on SuicideWatch, we look at the entire conversation ``thread''. In other words, we start with the content from the Original Poster (OP), but we also include the hierarchically nested thread of replies to the original post, the replies to those replies, and so on.
%
Further, most previous studies have aimed at studying the \emph{content} of posts and their characteristics in relation to other posts. But an important aspect of online communities is their supportive \emph{function}, where users can turn to these platforms not only to express their thoughts and concerns, but also to receive support from the community. This support often manifests as an emergent conversation between many users and the one in distress. Hence here we propose a framework that captures the structure of a conversation thread and develop metrics that capture the \emph{macroscopic} properties of a conversation that involve the entire thread and the users participating in it as well as \emph{mesocopic} properties of a conversation that involve the immediate interactions with the one in distress. 

\begin{figure*}[!ht]
    \centering
    \includegraphics[width=0.7\linewidth]{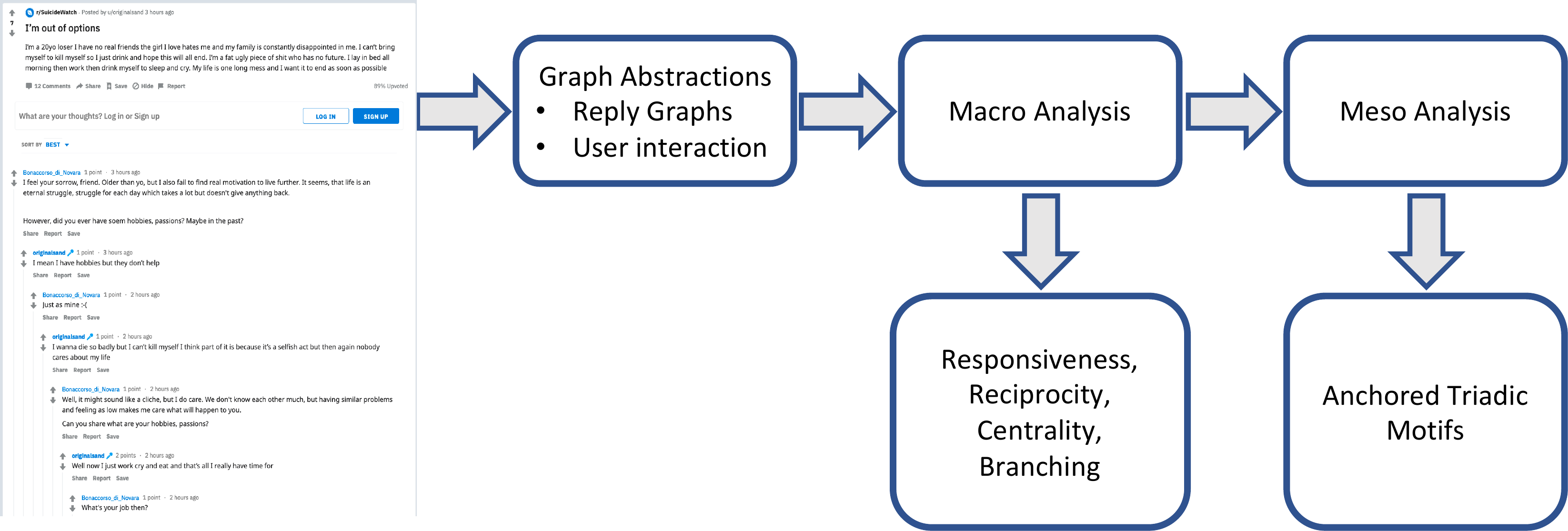}
    \caption{A reddit thread is converted into abstractions (Reply graphs and User interaction graphs). Macroscopic and mesoscopic analysis is performed on these graphs, and statistical over- or under- representation of these metrics is evaluated.}
    \label{fig:pipeline}
\end{figure*}

To model the conversation structure, we represent conversations in a forum using two graph-based abstractions: \textit{User interaction graphs}, which model the user-to-user exchange of messages, and  \textit{reply graphs}, which capture the structure of the dialogue on the forum, see example in Figure \ref{Fig:GraphExamples}. The complete processing pipeline can be seen in Figure~\ref{fig:pipeline}. We describe the pipeline and the metrics in detail in the \nameref{section:methods} section.
We then propose metrics that quantify the macroscopic structure of the two graphs we construct:
\textbf{Responsiveness} measures how quickly the responses accumulate in the reply graph; \textbf{Centrality} of the OP  measures how important the OP is to the conversation thread by computing the betweeness centrality\cite{white1994betweenness} of the OP in the user graph; \textbf{Reciprocity} measures the extent to which users obtain replies to their posts, by computing the fraction of edges in the user graph that are bidirectional; \textbf{Branching factor} measures how the reply graph fans out, i.e., the number of replies a post obtains. 

To measure local or mesoscopic structure, we turn to network motifs~\cite{milo2002network}. We propose a new method to count and characterise local structures, called anchored triadic motifs. Triadic motifs traditionally consider three nodes at a time~\cite{milo2002network}. Given the primacy of the OP, our method distinguishes variants based on where the OP is situated in a triad, to understand how the local patterns of  communications support the OP.  
In summary, this paper makes four key contributions: 

\begin{figure*}[!h]
    \centering
    \subfloat[]{
        \includegraphics[width=0.2\textwidth ]{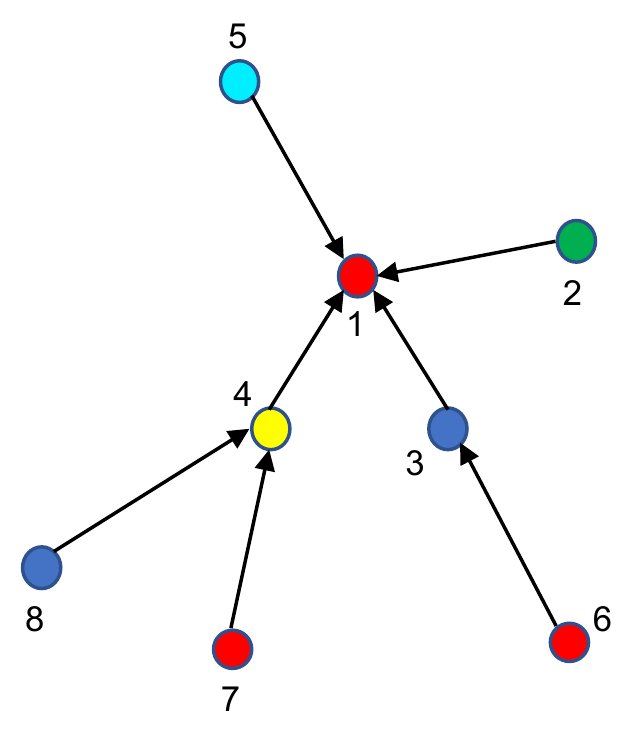}
        \label{fig:rGraphSW}
    }
\hspace{30mm}
    \subfloat[]{
        \includegraphics[width=0.2\textwidth]{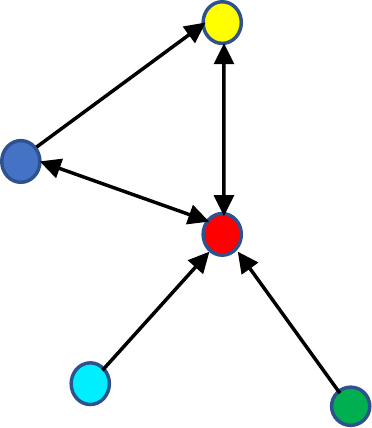}
        \label{fig:uGraphSW}
    }
    
    \caption{ Figure ~\ref{fig:rGraphSW} shows a sample reply graph constructed from a real thread in SW that contains 8 posts by 5 unique users. Each node represents a post and a directed edge is drawn from one node to another node when the first node is a reply to the second node. Thus, for example, Node 1 is the original post, with four replies (posts 2, 3, 4 and 5). Each node is given a colour based on the author of the post that the node represents, and each distinct colour represents a distinct author. Thus, from the reply graph, we can deduce that the original poster (Red node) obtained replies from the blue, green, yellow and purple users. In turn, the red node replied back to purple and yellow nodes ,but not to the blue and green nodes. The entire list of directed interactions is captured in a user interaction graph in Figure ~\ref{fig:uGraphSW}, where each coloured node represents the corresponding user who wrote a post on the thread, and the directed edges represent the replies.}
    \label{Fig:GraphExamples}
\end{figure*}

\setlist{nolistsep}
\begin{itemize}[noitemsep]
    \item We develop a framework  that abstracts out both the structure and semantics of a threaded conversation on the web
    \item Using this abstraction, we develop metrics which quantify the macroscopic (thread-wide) properties of conversations on SuicideWatch. 
    \item We develop a new method, which we term \textit{anchored triadic motifs}, to understand the mesoscopic or local structure of SuicideWatch conversations using triadic network motifs. Our method adapts triads by anchoring on the position of the Original Poster (OP), thereby distinguishing the OP from the other posters and helps understand how the conversation supports the OP's needs. 
    \item We show that there are significant statistical differences, both in the macroscopic and mesoscopic realm, that differentiate a SuicideWatch conversation from a generic conversation. 
\end{itemize}

\section{Results}

Reddit is a platform where a user can create a post or reply to a root post (RP) submitted by an original poster (OP) in a subreddit, and other reddit users can interact by posting at different levels of the thread, or by up or down voting posts. We analyzed RPs in the SuicideWatch subreddit (SW), building on the work of Gkotsis et al. \cite{gkotsis2017characterisation}.
We crawled SW to get entire conversation threads,
iteratively pursuing each conversation at progressively deeper levels of replies until the whole thread had been obtained. The code to crawl reddit for threads can be found at \textit{https://github.com/sagarjoglekar/redditTools}. This resulted in a dataset of over 50,754 SW threads totaling in 419,555 individual posts. 
To provide a baseline against which to compare nature of conversations on the SW sub-reddit, we acquired a similar number (49,773) of baseline threads from any other subreddit popular enough to land on the frontpage (FP). This resulted in a baseline dataset of 3,011,765 posts. Further details on how these were acquired are presented in the \nameref{section:methods} section. We compare the two corpora -- SW and FP -- at two scales: the first is a macroscopic analysis that considers features of entire threads; second we perform a mesoscopic analysis by considering local structural relations between nodes and their neighbours within user graphs corresponding to each thread. Our analysis finds several factors that distinguish SW conversations from FP conversations. 
\subsection{Macro Analysis of SW and FP Conversations}
 \begin{figure}[!h]
	\centering
	\subfloat[]{
		\includegraphics[width=0.25\textwidth ]{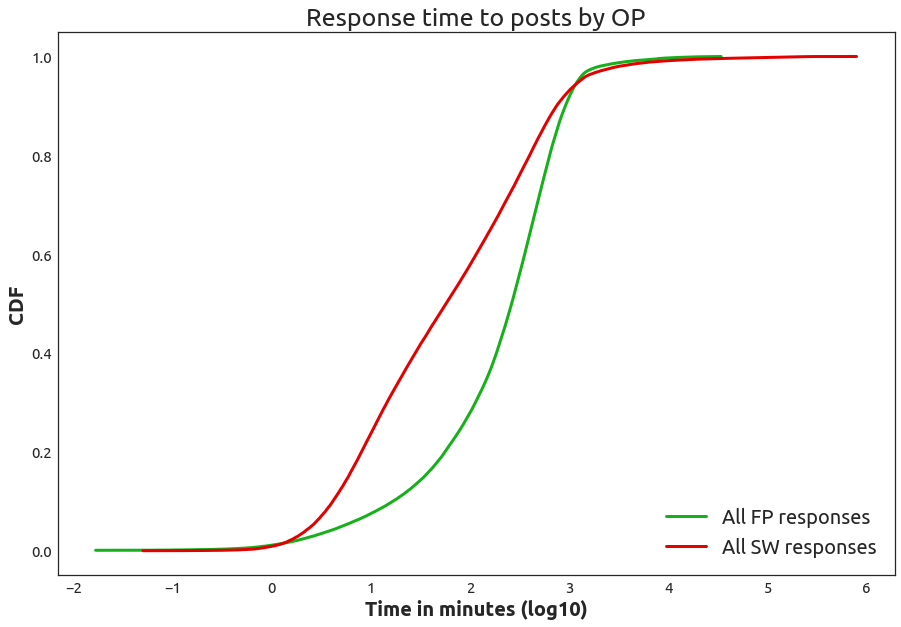}
		\label{fig:urgency}
	}
	\subfloat[]{
		\includegraphics[width=0.25\linewidth ]{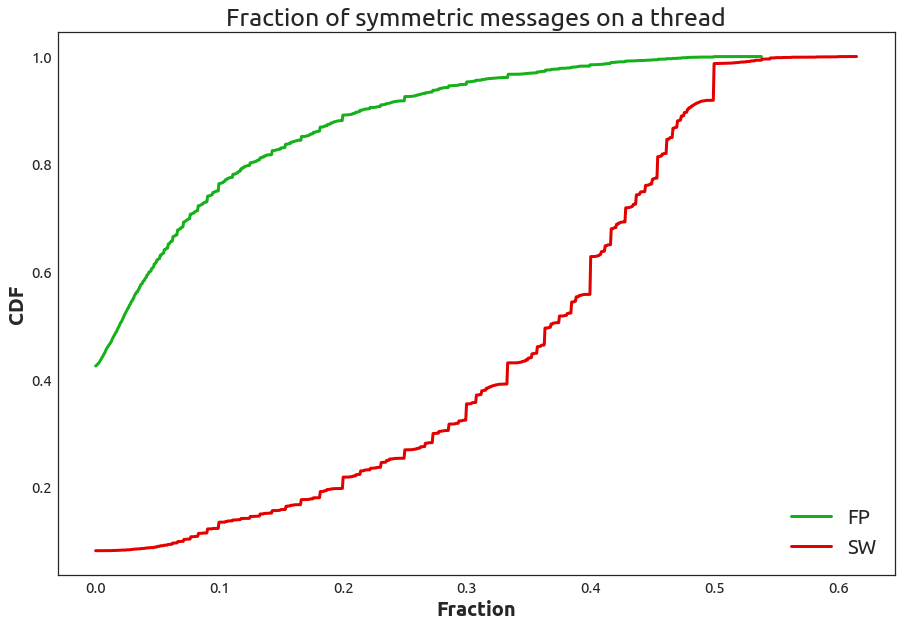}
		\label{fig:sym}
	}
    \subfloat[]{
		\includegraphics[width=0.25\linewidth ]{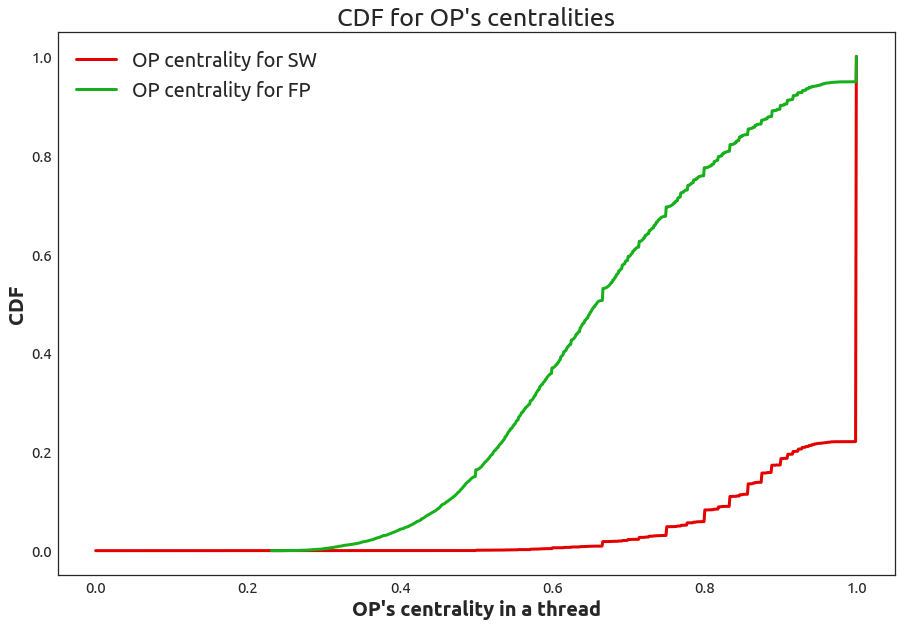}
		\label{fig:centrality}
	}
    \subfloat[]{
		\includegraphics[width=0.25\linewidth ]{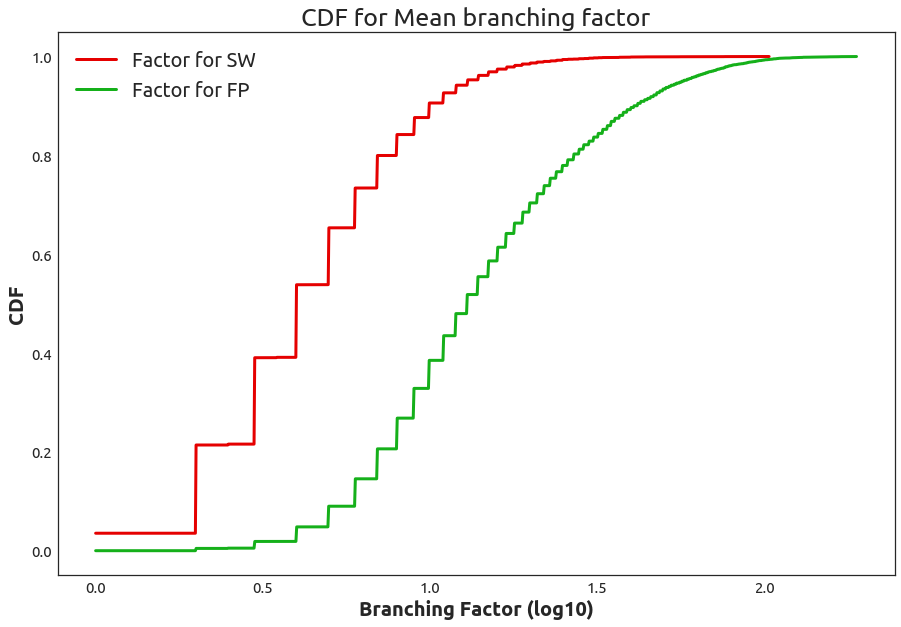}
		\label{fig:branching}
	}
	
	\subfloat[]{
        \includegraphics[width=0.25\linewidth ]{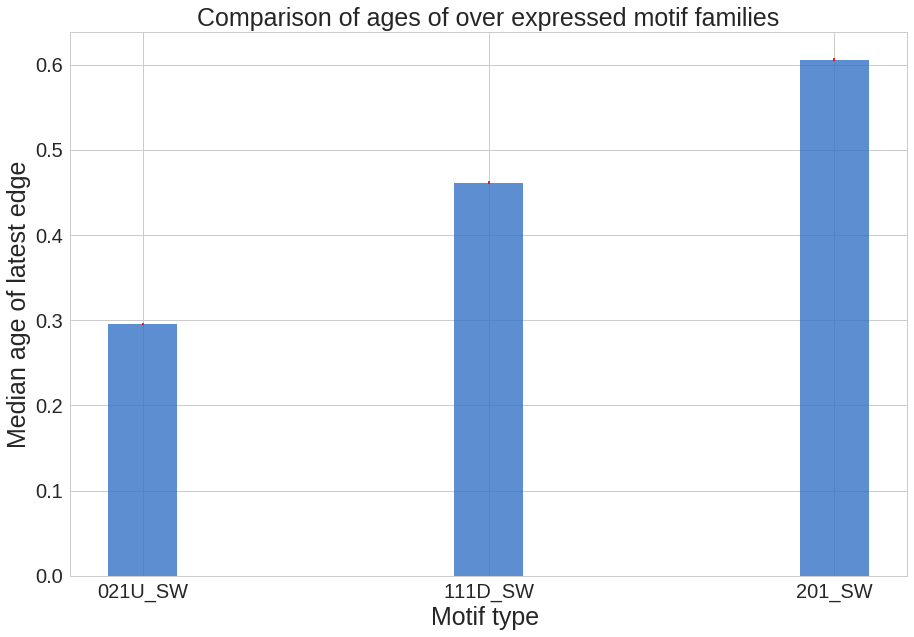}
        \label{fig:motif-progression}
    }

\caption{This panel shows the cumulative distribution  (CDFs) of Macroscopic features for SuicideWatch sub-reddit data (SW, in blue). These are compared with the control dataset of generic conversations on reddit from the FrontPage (FP, in green). \ref{fig:urgency} depicts results for Urgency; \ref{fig:sym} for Reciprocity; \ref{fig:centrality} for OP's  centrality in an interaction graph; 
and \ref{fig:branching} for Branching. SW conversations score higher on reciprocity, urgency and semantic alignment than FP. The SW conversations tend to branch less and tend to have higher centrality when compared to FP. Figure \ref{fig:motif-progression} represents the median completion times of the three motifs over expressed in SW, where the OP is at the apex (most central) position. This plot shows that as the time goes by, the symmetric nature of interaction between the OP and those who engage with them increases.}
\end{figure}

\subsubsection{Responsiveness: Users respond faster on $SW$ than on other subreddits}
To understand how responses on SW compared to other sub-reddit threads on FP, we calculate differences between the posting times between consecutive messages in a reply graph. The time that elapses between successive messages, i.e., the inter-message times, is taken as an indication of the urgency of how responsive a thread is. Figure \ref{fig:urgency} shows a comparison of inter-message response times for SW and FP threads, using the empirical Cumulative Distribution Function (CDF). Given a point $(x,y)$ in a CDF, $y$ should be interpreted as the frequentist probability that a particular variable (inter-message response times in this case) is less than the value $x$. Thus, Figure \ref{fig:urgency} shows that \textit{responses in SW tend to be much faster than in other sub-reddits}, suggesting that the community sees a need for urgency in responses. 

\subsubsection{Reciprocity: Interactions on SW are more likely to be bidirectional}
Next, we look at whether two users talk with each other -- i.e., if user \textbf{A} replies to a user \textbf{B}, does \textbf{B} reply back to \textbf{A}?  Figure \ref{fig:sym} plots the empirical CDF of the fraction of posts which are reciprocated, showing a vast difference -- \textit{conversations on  SW are much more reciprocal than other subreddits}: the \textit{median} value for $U_{sym}$ for SW is 50\% whereas for FP is 2.6\%. 

\subsubsection{Centrality: $OP$ is more central in SW conversations}
To understand whether and to what extent SW conversations revolve around the $OP$ (who may have posted in distress), we consider the user interaction graph of each thread, and plot the \textit{betweenness centrality} of the \textit{OP}. Betweenness centrality of a node measures how often that node is on the shortest path connecting two other nodes, and as such, is a measure of how central the node is in the graph. Figure~\ref{fig:centrality} shows the empirical CDFs of centralities. It can be seen that the \textit{$OP$ has much higher betweenness centrality in SW conversations than FP conversations}. 



\subsubsection{Branching: SW conversations branch out considerably less compared to FP}
We next measure the \textit{number} of responses or branching factor of the reply graph, using the formula described in Section \nameref{sec:branching}. Figure \ref{fig:branching} shows that SW threads branch out considerably less than threads on other subreddits, which could be indicative of the first few replies satisfying the need for response embedded in the posts they are replying to (e.g., if the  post is a query, the initial replies could be providing all the information asked; if the post is a call for help; the initial replies could be providing the necessary level of support). 


\subsection{Mesoscopic analysis: Patterns in local interactions}

\begin{figure}
	\centering
    \subfloat[]{
    	\includegraphics[width=0.8\textwidth]{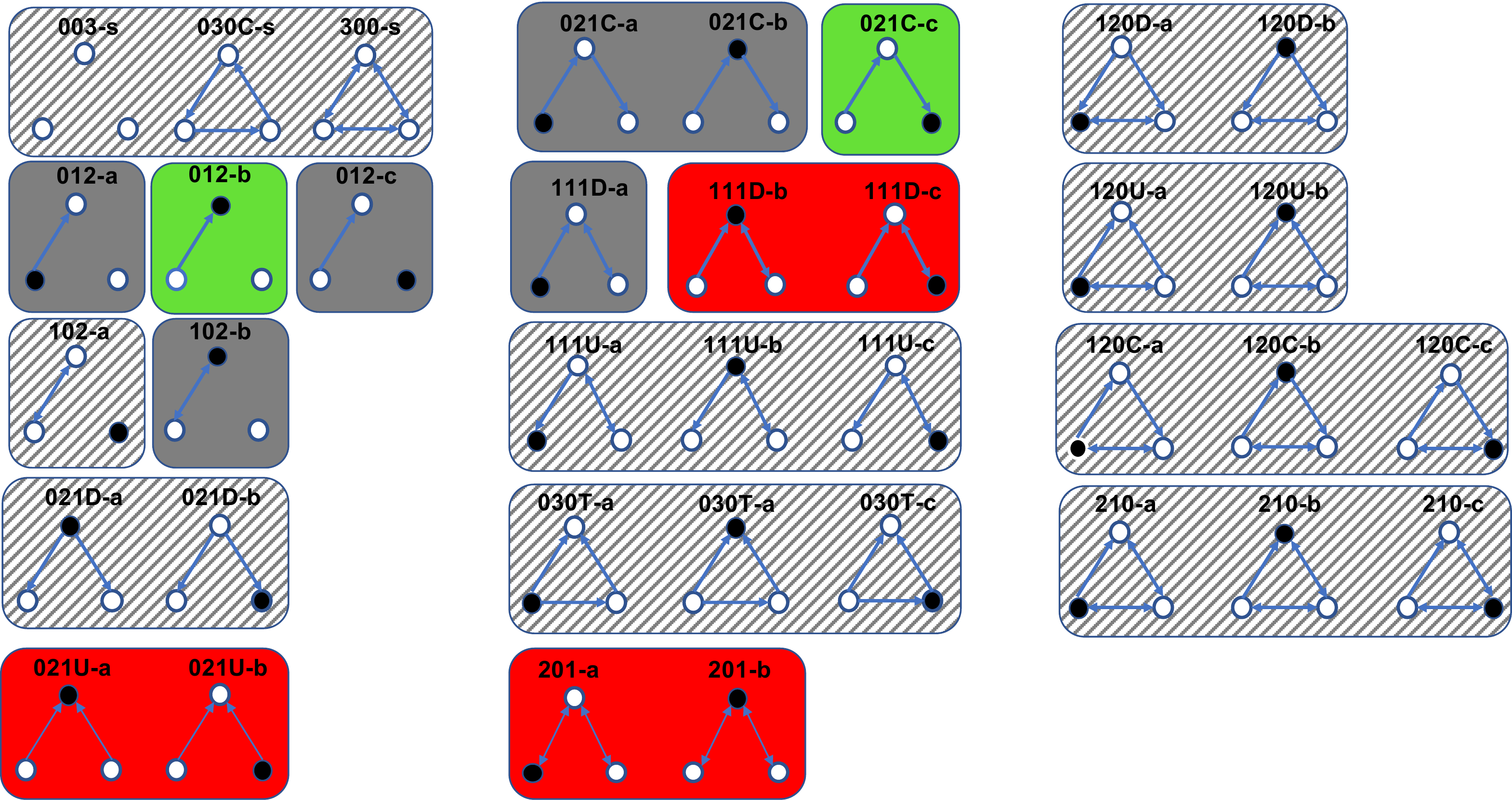}
    	\label{fig:motifs}
	}
	
	\subfloat[]{
        \includegraphics[width=0.2\linewidth ]{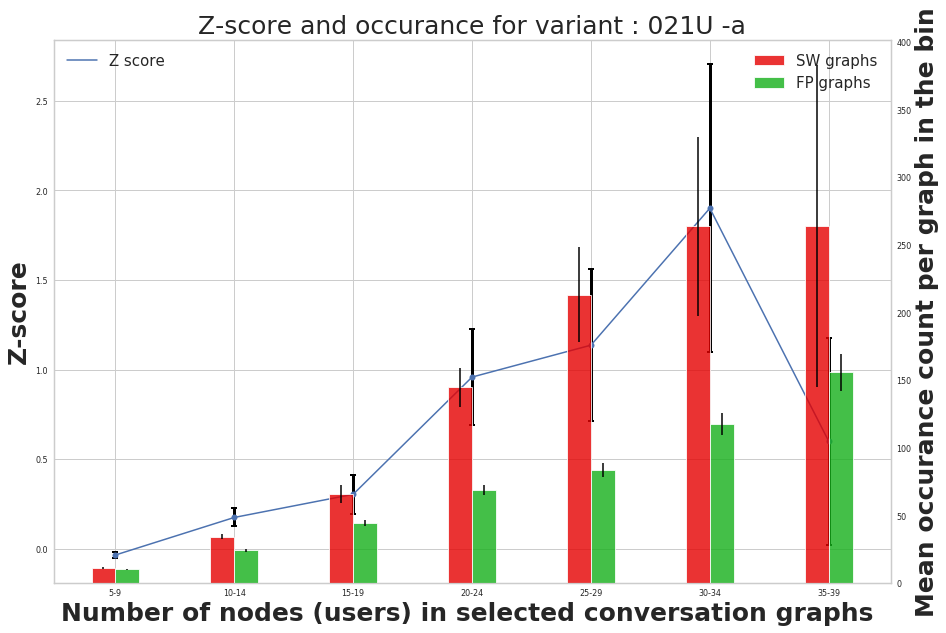}
        \label{fig:021U-a}
    }
    \subfloat[]{
        \includegraphics[width=0.2\linewidth ]{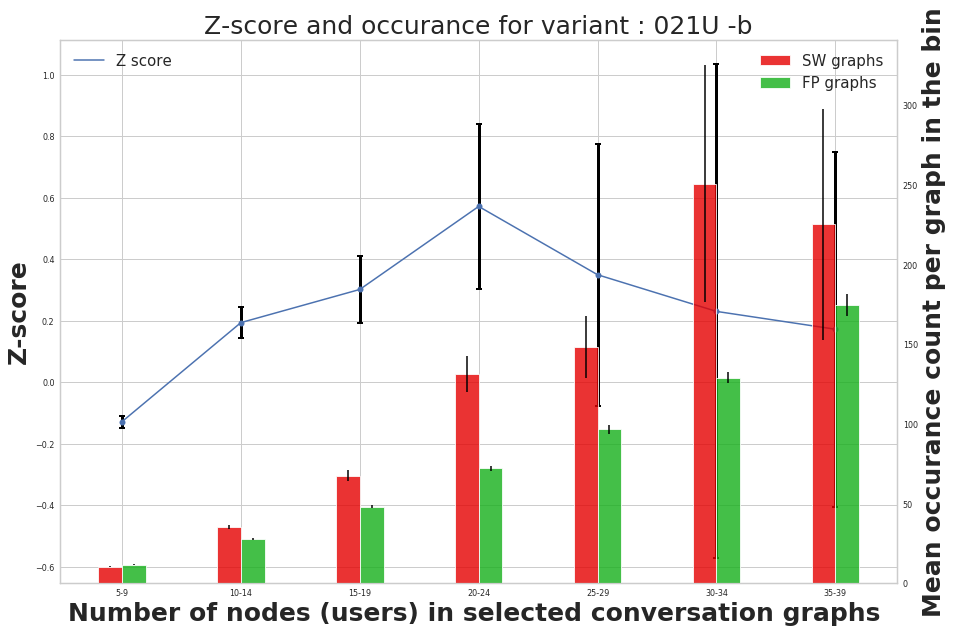}
        \label{fig:021U-b}
    }
    \subfloat[]{
        \includegraphics[width=0.2\linewidth ]{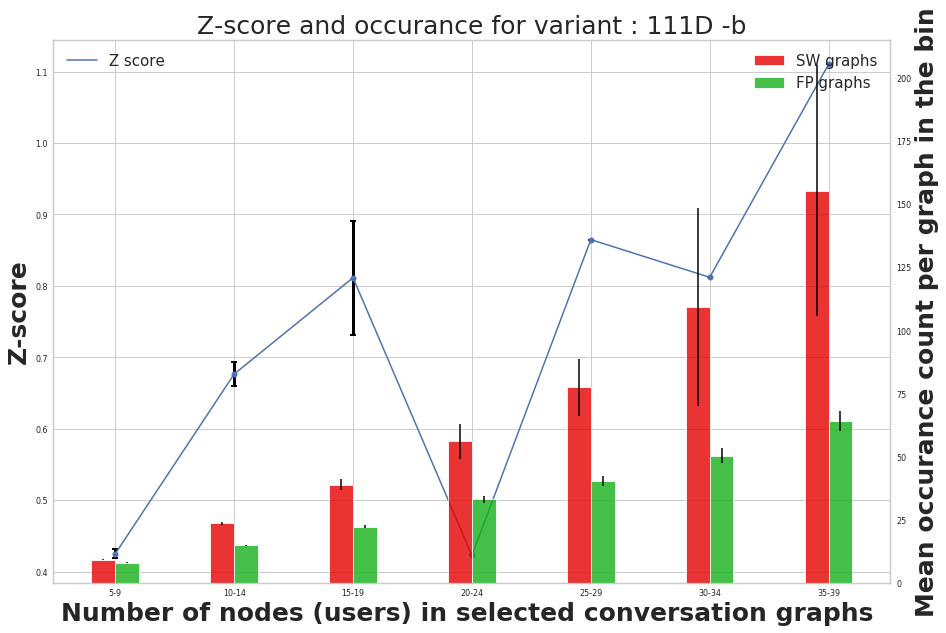}
        \label{fig:111D-b}
    }
    \subfloat[]{
        \includegraphics[width=0.2\linewidth ]{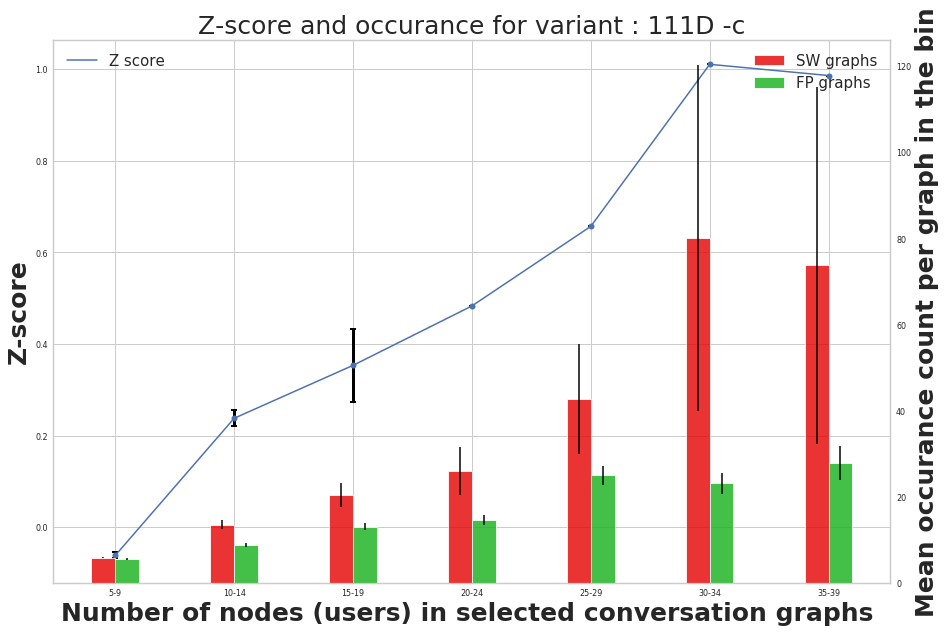}
        \label{fig:111D-c}
    }   
    
    \subfloat[]{
		\includegraphics[width=0.2\linewidth ]{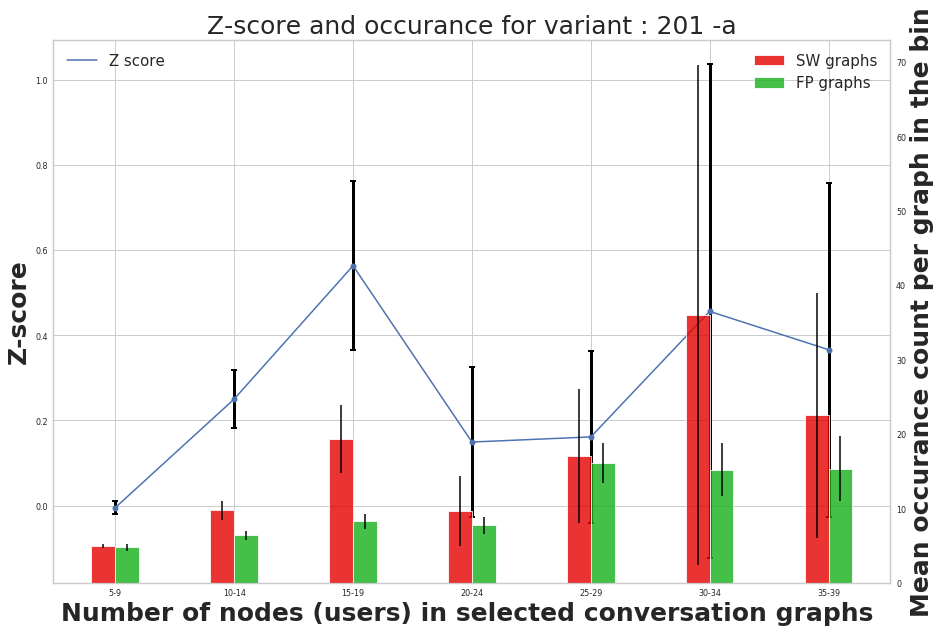}
		\label{fig:201-a}
	}
    \subfloat[]{
		\includegraphics[width=0.2\linewidth ]{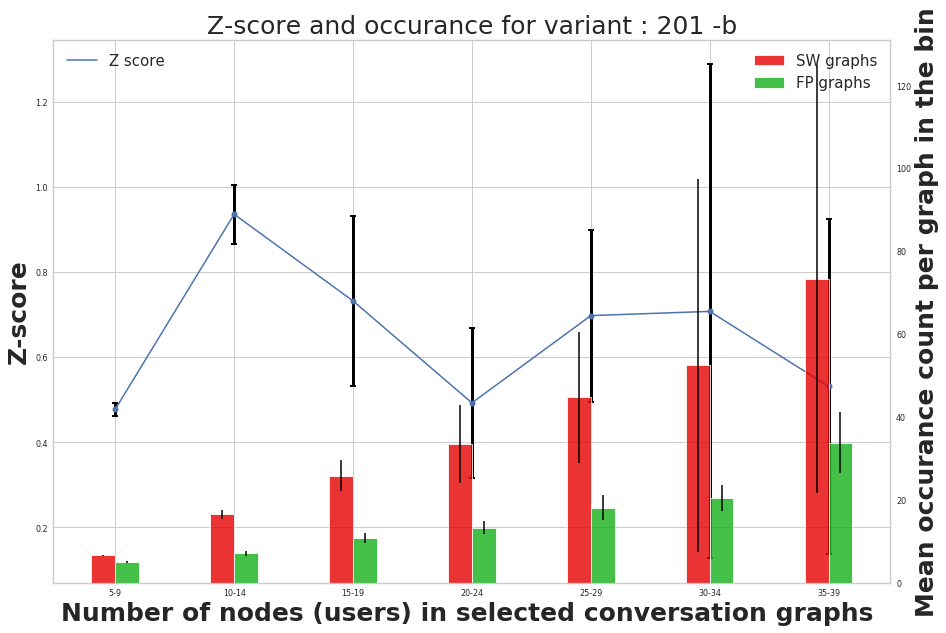}
		\label{fig:201-b}
	}
    \subfloat[]{
		\includegraphics[width=0.2\linewidth ]{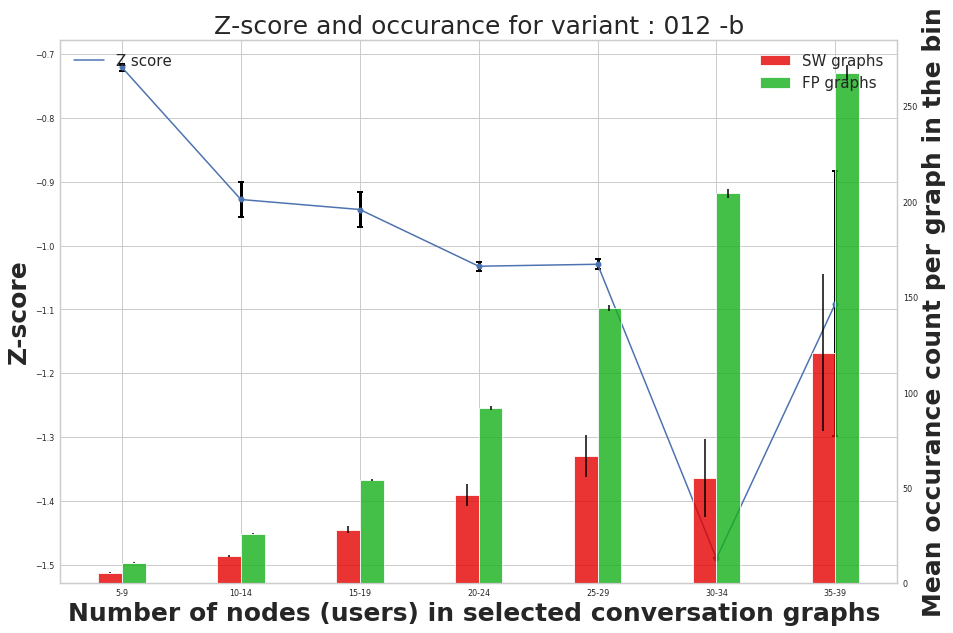}
		\label{fig:012-b}
	}	
    \subfloat[]{
        \includegraphics[width=0.2\linewidth ]{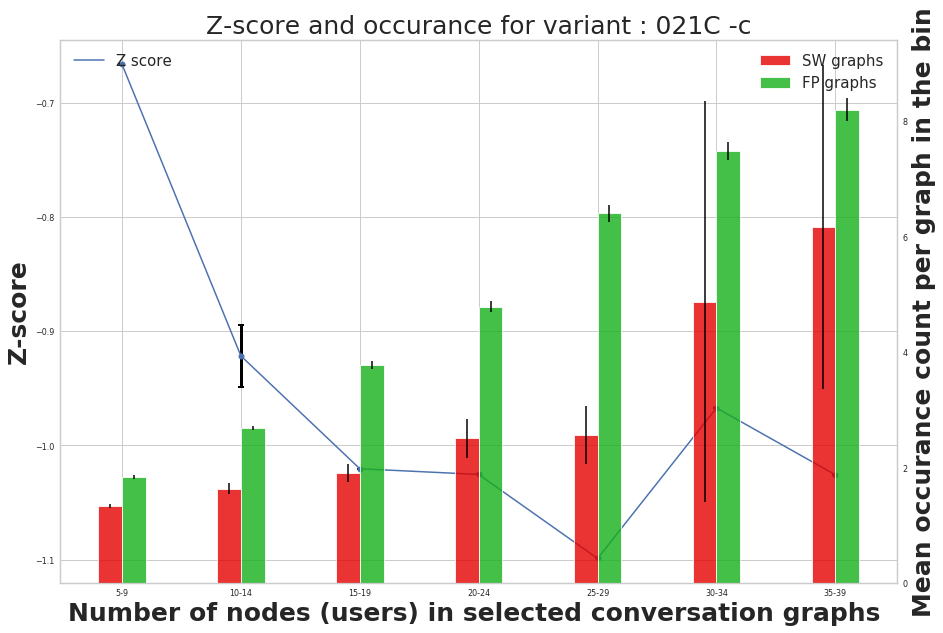}
        \label{fig:021C-c}
    }
    
\caption{
Figure \ref{fig:motifs} shows the 36 different types of Anchored Triadic motifs which are statistically compared between FP and SW graphs. The motifs with \textbf{green boxes} are \textbf{over expressed in the baseline dataset (FP)} by a significant amount. The motifs with \textbf{red boxes} are \textbf{over-expressed in the SuicideWatch (SW) dataset} by significant amount. The motifs with \textbf{grey boxes} are present in significant numbers in both datasets, but neither over nor under expressed in any datasets based on their Z scores. The motifs in \textbf{grey hatched boxes} are very rare in both the baseline and suicide watch datasets, with less than 5 mean occurrences per graph per bin.
Figures \ref{fig:021U-a}--
\ref{fig:012-b}  and \ref{fig:021C-c} show the side by side comparison of motif occurrences for SW and FP across different bins for motifs that are either over or underexpressed (i.e., coloured green or red in Fig.~\ref{fig:motifs}. The Z-score from the comparison is plotted as a blue trace, alongside the mean population of the motif in both SW and FP in a selected bin. For completeness, we report the results for other remaining motifs in the supplementary material.}
\label{Fig:motif_expressed}
\end{figure}

It is often useful to express large interaction graphs as the sum of local interactions between two or three nodes at a time. This method is 
prevalent in the Social Sciences, for studying social structures by looking at local interactions between agents\cite{faust20077}. Such analysis is also useful in expressing local structures in large graphs and has been used in several network analysis works\cite{wang2014triadic,shizuka2015network}.
For this reason we conduct a census of the 36 Anchored triadic motifs (see Figure \ref{fig:motifs}), using methods further described in Section \nameref{sec:motif}), across all the selected graphs. Anchored motifs extend the concept of triadic network motifs by distinguishing different variants based on the position of a special node, which we take here to be the Original Poster (OP) who started the thread. By distinguishing the OP's position, we are able to reason about how a particular motif may help serve the needs of the OP. Commonly, motif analysis compares the occurrence of each triad in a real network against a baseline, for instance a null model created using generative processes (e.g. random graphs). In this case, we compare the motifs seen in SuicideWatch against the set of all graphs that belong to generic conversations from the Frontpage (FP).  
We perform binning of user graphs as described in Section \ref{sec:motif}, and perform over- or under-expression analysis in comparison with motif census performed on FP as the baseline null model. We use Z-scores of the motif occurrences as a metric to measure statistical significance. We are interested in anchored motifs which are present in significant numbers as well as have strong over or under expression. We classify a motif population as significant if the mean motif population goes above 10 for any of the 7 bins. We consider a motif over/under expressed if the Z-score is either greater than 1 or less than -1 for at-least 1 bin. A motif which has significant mean population but has a Z-score between -1 and 1 is considered equally expressed. Figure \ref{Fig:motif_expressed} shows all the 8 motifs which are statistically significant and over/under expressed. 

We find that anchored motif variants \textbf{021U-a, 021U-b, 111D-b, 111D-c, 201-a and 201-b} are significantly over-expressed in SW conversations across all sizes of graphs as seen from figures \ref{fig:021U-a},\ref{fig:021U-b},\ref{fig:111D-b},\ref{fig:111D-c}, \ref{fig:201-a},\ref{fig:201-b}. Similarly anchored motif variants \textbf{012-b and 021C-c} are significantly over-expressed in the null model (FP) graphs across all sizes. 

We look at the median completion times for 3 of the 5 over expressed motifs (021U-a, 111D-b and 201-b), by plotting the median age of the last established edge in the motif as a fraction of the entire lifetime of the thread (Figure \ref{fig:motif-progression}). These three motifs share a peculiar property in that they all have the OP at the apex (most central) position. We observe that as the time goes by, the symmetric edges between the OP and those who engage with them increases.  

From previous studies on triadic structure, it was inferred that transitive triads are naturally more common than expected in social structures of apes and humans~\cite{shizuka2015network}. Interestingly, our analysis shows that transitive triads are rarer in SW, as compared with the FP conversations. 
These patterns in local interactions indicate that conversations in SW tend to be more $OP$ centric, with non-transitive dialogues between the $OP$ and users who respond to their calls for help. As a consequence, the $OP$ tends to be highly central in the conversation as well as part of several mutual interactions. These behaviours are unique to SW, i.e., observed more in SW than when compared with conversations on other subreddits (FP).

\section{Discussion}

The results show that there are several factors which distinguish a a SuicideWatch (SW) conversation from a comparative baseline of  conversations gathered from the front page (FP) of Reddit. Based purely on the \textit{ structure} of the conversations, we identified four clear differences between the macroscopic features of SW and FP: 
The speed or \textit{urgency of response to the OP in SW is faster than for FP}, which is as would be intuitively expected of a subreddit set up as a place of support for “vulnerable OPs”.  
Just as in group therapy, it is individual clients and their larger relationship within therapy that is the agent of change \cite{yalom_theory_1995}, and the same is reflected in this peer support forum.   
Features of relational communication \cite{rogers_overview_1983} are that \textit{SW shows more symmetry and reciprocity than FP}, and the \textit{OP is central to the communication}. Studying the interlocking and reciprocal effects of each interactor on the other has been key to understanding “therapy as a system” in face-to-face therapeutic encounters also \cite{de_shazer_putting_1991}.
What is radically different from a clinical context is that the posters are not healthcare seeking and may be on SW precisely because they are seeking alternative support and there is no ‘professional’ facilitating the discussion.  The closest analogy in a live group setting is “fishbowls” (used in certain group counselling courses) \cite{keim_groupwork_2013} where there is an inner ring of discussants (the OP and other posters on the thread) who are observed by an outer-ring of observers (in SW, a parallel may be drawn with the moderators who manually examine comments and delete those that threaten or violate the thread’s specified codes
and ban trolls 
\cite{choudhury_language_2017}).


In communication accommodation theory (CAT)\cite{coupland_introduction_1988}, which was developed for face-to-face conversations between two people (a dyad) but has now been extended to mediated dyadic discussions (e.g. on Twitter) without temporal immediacy \cite{lipinski-harten_comparison_2012}, the concept of accommodation has two opposite forms: convergence and divergence.  Convergence is mimicking of the conversational partner's style and divergence is avoidance of the style. This phenomenon may be reflected in SW by less branching or digression in the conversation thread compared to FP.

At a mesoscopic level, the most striking features of the anchored triadic motifs which occur in statistically significant numbers (shown with a solid, non hatched background in Fig. \ref{fig:motifs}) are that none of them involve all three nodes, suggesting that dyadic communications (e.g., providing an answer to a question) are the primary focus.  Of course it would be extrapolating to assume this is supportive communication, and there would need to be further qualitative research into the \textit{content} of the threads that demonstrated these motifs.  Similarly when considering the anchored triadic motifs which are under expressed in both FP and SW datasets (grey hatched in Fig \ref{fig:motifs}), it is worth noting that although these are statistically rare in the current work, they could be worth exploring in other datasets.

Both of the anchored triadic motifs that are over expressed in the baseline FP conversations and by comparison are under expressed in SW (motifs that are shown in green in Fig. \ref{fig:motifs}), show non-conversational, non-reciprocal patterns of serial communications between respondents to an OP (021C-c) and unidirectional response to an OP from one respondent (012-b). In contrast, those over expressed in SW (shown in red) have two arrow heads pointing towards the apex nodes, suggesting that communication is directed towards one participant. Except for 021U-a and -b, the other motifs over expressed in SW all have at least one bidirectional conversation, reflecting the high levels of reciprocity in SW. 

Internet health forums have been studied in several instances and their utility has been shown to be of value in cases of chronic illnesses\cite{Joglekar2018}, addictions\cite{wood2009evaluation} and mental health issues \cite{gkotsis2017characterisation,de2013social}. However, most of these studies have focused on quantitatively analysing
the content discussed and the linguistic signatures of how these communities interact.
Here, we have instead focused on developing ways to quantitatively analyse the \emph{structure} of online communication, and study how and whether this structure reveals patterns of peer and community support.
To that end, this is the first attempt at finding topological discriminatory factors between supportive and generic conversations on social media forums. Our focus on structure rather than content means that our methods can potentially be extended to other languages more easily.

The public health implications of this work are that the distinctive supportive network structures and the content of their posts should be studied in more detail to investigate what works well and why.  This could help educate peer moderators to have a better overview of the subreddits they moderate and the ongoing conversation. Topological features could be used in addition to the community signals they already use, such as numbers of upvotes or downvotes, or referring to comments flagged by community members  \cite{choudhury_language_2017}.  Similarly, studying the less supportive motifs could lead to insights into \emph{why} certain interactions are unhelpful, and might allow automated detection of such interactions so that moderators are able to moderate such comments in a more timely fashion.  The results obtained could also be used as a selection strategy for purposefully sampling more supportive networks. We believe that the novel framework for macro and meso analysis of supportive online communities we present here can provide important directions for future research in this area.  
\section{Methods}
\label{section:methods}
\subsection{Background on Reddit conversations}
Building on the brief descriptions in the previous sections, here we provide a more detailed background of Reddit conversations: In most forum based platforms such as reddit,  users interact in a nested dialogue fashion, where an Original Poster or $OP$ posts content called a Root Post or $RP$ to start  a new discussion thread. This thread is then open for comments by all the community users. In case of Reddit, such a community is called a Subreddit. Subreddits like SuicideWatch consist of a moderated collection of posts from users who subscribe to that community or subreddit. These users may post new threads onto the subreddit as long as the post follows the subreddit rules. Enforcement of these rules is the responsibility of the moderators. 

\subsection{Datasets} \label{sec:data}
\begin{figure}
    \centering
    \subfloat[]{
        \includegraphics[width=0.25\textwidth ]{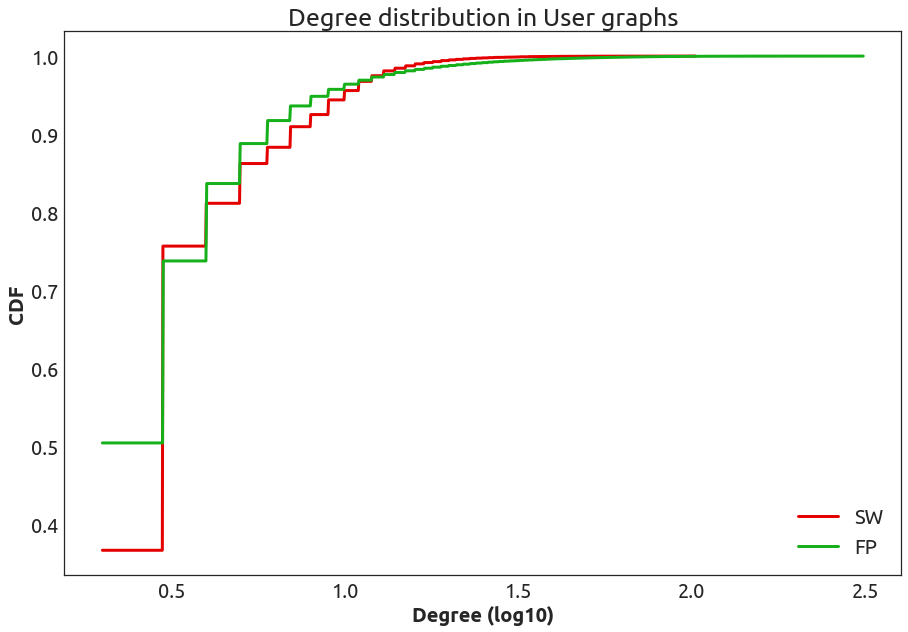}
        \label{fig:degUgraph}
    }
    \subfloat[]{
        \includegraphics[width=0.25\linewidth]{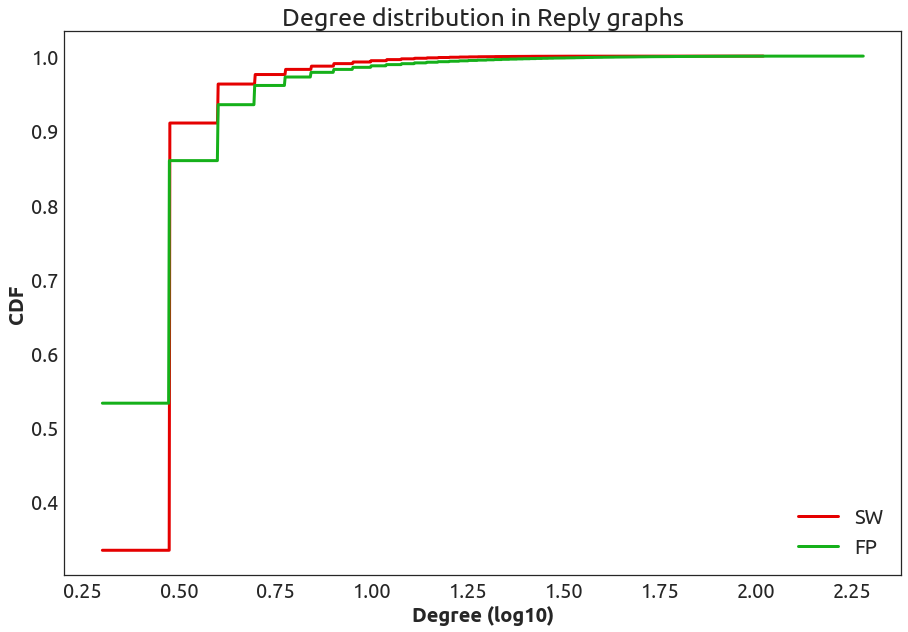}
        \label{fig:degRgraph}
    }\caption{Degree distributions  (a) User Graph and (b) Reply Graphs in FP and SW, showing that the two datasets are comparable.}
    \label{fig:degdist}
\end{figure}

The focus our work is the subreddit r/SuicideWatch. We study this using a seed dataset~\cite{gkotsis2017characterisation}, that consists of root posts from the subreddit r/SuicideWatch. Building on this dataset, we acquire the entire thread structure of all the root posts in the data by recursively obtaining all replies to the root posts, replies to those replies and so on, until we reach posts which do not have any further replies. This results in the acquisition of 50,754 threads from SuicideWatch (SW). To obtain a baseline of similar size for comparison purposes, we crawl the entire conversation threads of posts that appear on the front page of Reddit.com  for 2 weeks,  accumulating a second corpus (FP) of 49,773 reddit threads in the process. 
The two conversation datasets from r/SuicideWatch and Frontpage are very similar in terms of common summary statistics such as Degree distributions (see Fig.~\ref{fig:degdist}). 
Owing to the long tailed nature of the datasets, we perform our analysis on threads  which have at least 5 posts in addition to the root post.
We further clean the data, by removing threads where the root author has deleted their user account, which is a common practice to preserve anonymity in more controversial posts. The resulting dataset has 10,527 threads in SW and 11,070 threads in the baseline (FP). 

\subsection{Abstractions} \label{sec:abstractions}
To understand the dynamics of supportive conversations, we develop two abstractions: 
\subsubsection{Reply Graphs} \label{sec:reply_graphs}
To mimic
the structure of conversation threads on Reddit 
we define
an abstraction 
that
we term as \textit{reply graphs}, and denote as $R = (P,E)$. The nodes $P$ consist of all the posts in the thread. The root post is labeled as $p_0 \in P$ and $k^{th}$ post in chronological order as $p_k \in P$. A directed edge ($p_i,p_j) \in E$ is drawn from $p_i$ to $p_j$ if $p_j$ is a reply to $p_i$. \ref{Fig:GraphExamples} (a) presents an example. Note that in platforms such as Reddit, where each response can only reply to one other post, reply graphs end up being reply \textit{trees}.

\subsubsection{User interaction Graphs} \label{sec:interaction_graphs}
The second abstraction, which we term \textit{user interaction graphs},  represents each thread as a directed graph $G = (V,E)$ where $V$ is the set of all users participating in a particular thread and a directed edge $(v_i, v_j) \in E$ is drawn between two users $v_i$ and $v_j$ if user $v_i$ responds to a post by user $v_j$. \ref{Fig:GraphExamples} (b) presents an example. Note that unlike reply trees, user interaction graphs of Reddit conversations can be full graphs, and may include cycles. 

\subsection{Macroscopic graph metrics}
The abstractions are used to extract the following structural metrics from the conversation threads. These metrics are then used to validate structural differences between supportive conversations and generic casual conversations from our baseline set.

\subsubsection{Responsiveness}
To understand the speed with which users in a subreddit (whether in SW or in other subreddits as represented in FP) respond to the $OP$ and each other,  we calculate differences between the posting times between consecutive response messages in a reply graph. We then compute the median response times per thread.

\subsubsection{Reciprocity}
Reciprocity measures the extent to which users' posts obtain responses, and is measured as the fraction of edges in the user graph that are bidirectional:
$$
U_{sym}=\frac{\textit{total number of bidirectional edges in a user graph}}{\textit{Total number of edges in the graph}}
$$

\subsubsection{OP Centrality} \label{sec:centrality}
Node centrality is a metric that measures how central a node is in a network. We study how central the OP is to a conversation thread by computing how often it lies on the shortest connecting paths between other pairs of nodes in the user interaction graph. Betweenness centrality is formally defined as 
$$
g(OP) = \sum_{s \neq v \neq t}\frac{\sigma_{st}(OP)}{\sigma_{st}}
$$
where $\sigma_{st}$ is the total number of shortest paths from node $s$ to node $t$; 
$\sigma_{st}(OP)$ is the number of 
paths that pass through $OP$. 


\subsubsection{Branching Factor}\label{sec:branching}
Branching factor is a metric that reflects the fanning out of a conversation as it evolves. To measure this on Reddit reply trees. we compute the average number of replies obtained by each post, i.e., the average in-degree of nodes in each reply graph. 

\subsection{Mesoscopic graph metrics: Anchored triadic motifs  } \label{sec:motif}
Network motifs are local sub-networks, typically of 2 or 3 nodes which are connected together. Such local patterns are highly useful in quantifying local interactions and the resulting macro structure of the network\cite{milo2002network}. They have been used in a variety of applications, from economics \cite{zhang2014dynamic} to cellular protein-protein interaction networks \cite{yeger2004network}. These local interaction patterns have been fundamental in the study of social structural processes\cite{faust20077}. They help social scientists quantify the type of hierarchies in the social network\cite{davis1967clustering,davis1967structure}. Hence, we turn to network motifs to characterize the local structure of the converstion threads. 
However, SW conversations shows clear distinctions between the users who respond to a call for help and the user/s who are asking for help (the $OP$). To accommodate this, we extend the concept of triadic motifs to create different variants of the same motif when the $OP$ is in different positions.

In conventional literature, the local interactions are measured through a census of 16 triadic motif patterns\cite{faust20077}, which cover all possible patterns of non-isomorphic triads which cannot be mapped or morphed into each other. In this method, there is no special treatment to any node, and position of all nodes is treated equally. To this, we introduce the notion of \textit{anchors}, or nodes with special importance, which in our case is the $OP$, the user who makes the initial post in the thread under consideration. 
By fixing a role for a node in a motif, each of the 16 triadic motifs as seen and developed in the field\cite{faust20077, holland1977method}, can be unravelled into 36 sub-variants of these motifs by varying the anchored node, as seen in Fig \ref{fig:motifs}. Each sub-variant is different from the other from the perspective of the anchored node. Some motifs yield three variants for each of the three positions that the $OP$ can be in. However, other motifs yield fewer variants, since two or more of the variants can be iso-morphic to each other even when the position of the $OP$ is distinguished.  Bataglej et.al's work\cite{Batagelj2001} developed a method for counting network motifs. We build on this and develop an efficient method for counting anchored network motifs. 
Each motif as seen in Figure \ref{fig:motifs} is named using the naming scheme  developed by Holland and Leinhardt\cite{holland1974statistical}. The first three numbers, follow a M-A-N pattern which signifies the number of "\textbf{M}utual" , "\textbf{A}symmetric" or "\textbf{N}ull" edges present in that particular triad. For example, the motif 030 has 0-Mutual(bi-directional), 3-Asymmetric(unidirectional) and 0-Null (disconnected) edges. There are some motifs with an added modifier letter (C-U-D-T) attached to further differentiate between different triad types with the same M-A-N pattern.
To this, we additionally attach a variant label (a, b or c) to distinguish the different anchored network motifs that result from the different positions of the $OP$.

To systematically understand the over or under expression of these anchored triadic motifs in the suicide watch community (SW), we use the user interaction graphs for the Front page (FP) baseline posts as a null model. We analyse 10,527 user interaction graphs from SW and 11,070 graphs from FP dataset.
We progressively select graphs with different sizes, i.e., graphs with differing numbers of users present in the interaction graphs. We bin both the FP and SW user interaction graphs as follows, based on the number of nodes interacting within a thread: 1 -- 5, 6 -- 10 , 11 -- 15 , 16 -- 20 , 21 -- 25, 26 -- 30, 31 -- 35 and 36 -- 40. The number of conversations that contain more than 40 unique users participating in the same conversation thread is extremely small in both SW and FP; hence we stop binning at this point.
Within each bin, we then perform a census, counting the number of occurrences of different anchored network motifs. 
Once the census is done, we calculate $Z_{scores}$ for the Suicide watch conversations, using FP conversations as the null model, to understand whether a given anchored network motif is over or under expressed in SW in relation to FP.

 
 We call the set of FrontPage and SuicideWatch graphs that belong to bin $b$ as $G^b_{FP}$ and $G^b_{SW}$ respectively. For a selected bin $b$, let $M$ graphs from $FP$ belong to $b$ and $N$ from $SW$ belong to $b$. We conduct the anchored motif census of the 36 motifs for both $G^b_{FP}$ and $G^b_{SW}$.  
To compute the null model, we require the mean ($\mu_{null}$) and standard deviation ($\sigma_{null}$) of the frequency distributions of all the 36 motifs found in the $G^b_{FP}$ graphs. This means we will have 36 values of ($\mu_{null}$) and ($\sigma_{null}$); one for each motif. We also compute the mean ($\mu_{SW}$) and standard deviation ($\sigma_{SW}$) for $SW$ dataset, and plot the means of $FP$ and $SW$ side by side as a comparison. The error bars represent standard errors ( $ e_{null} = \frac{\sigma_{null}}{\sqrt{M}}$ and $ e_{SW} = \frac{\sigma_{SW}}{\sqrt{N}}$). Plots of these mean frequencies for both datasets can be found in Figures \ref{fig:021U-a} -- \ref{fig:021C-c}.

Once we have the null model figures for the bin $b$ from $G^b_{FP}$ graphs, we  compare these with the $N$  graphs ($G^b_{SW}$) in order to compute the $Z_{score}$. For the $i^{th}$ motif, the score $Z_{i}$ is defined as 

$$Z_i = \frac{1}{N} \sum_{k=1}^{N} \frac{m_k^{SW} - \mu_{null}}{\sigma_{null}}$$ 

where $m_k^{SW}$ is the total number of the $i^{th}$ motif found in the $k^{th}$ graph in $G^b_{SW}$. We compute this $Z_{score}$ for all the 36 motifs across all the 7 bins. The trends in the value of this $Z_{score}$ are also plotted in Figure \ref{fig:motifs}. We consider a motif population as significant if the mean motif population goes above 10 for any of the 7 bins. We consider a motif over/under expressed if the Z-score is either greater than 1 or less than -1 for at least 1 bin. A motif which has significant mean population but has a Z-score between -1 and 1 is considered equally expressed. Figure \ref{fig:motifs} shows all the 8 motifs which are significant and over/under expressed, 
where as the measurements of the statistically insignificant motifs are included in the supplementary material.

\section{Data and Code Availability}
The datasets generated and analysed during the current study are available from the corresponding author upon reasonable request. The code for crawling the Reddit conversation structure, conducting census for Anchored Triadic Motifs, and analysis of the data can be found at \url{https://github.com/sagarjoglekar/redditTools}

\bibliography{replyGraphs}

\section{Contributions}
S.J. and N.S. designed the study. S.J. conducted the experiments. S.V., R.D. and N.S. helped with the conceptualization of the paper and interpretation of the results. S.J., S.V., R.D. and N.S. contributed to the writing of the manuscript and approved the final version of the manuscript.
\end{document}